\documentclass[aps,amssymb,amsmath,
twocolumn,showpacs,superscriptaddress,
groupedaddress]{revtex4-1}

\usepackage{graphicx}
\usepackage{dcolumn}
\usepackage{bm}
\usepackage[table,xcdraw]{xcolor}
\usepackage{mathrsfs}
\begin{document}
\title{Analytical evaluation of relativistic molecular integrals.\\
III. Computation and results for molecular auxiliary functions}
\author{A. Ba{\u g}c{\i}}
\email{abagci@pau.edu.tr}
\affiliation{Department of Physics, Faculty of Arts and Sciences, Pamukkale University, {\c C}amlaralt{\i}, K{\i}n{\i}kl{\i} Campus, 20160 Denizli, Turkey}
\author{P. E. Hoggan}
\affiliation{Institute Pascal, UMR 6602 CNRS, University Blaise Pascal, 24 avenue des Landais BP 80026, 63177 Aubiere Cedex, France}
\begin{abstract}
This work describes the fully analytical method for calculation of the molecular integrals over Slater$-$type orbitals with non$-$integer principal quantum numbers. These integrals are expressed through relativistic molecular auxiliary functions derived in our previous paper [Phys. Rev. E 91, 023303 (2015)]. The procedure for computation of the molecular auxiliary functions is detailed. It applies both in relativistic and non-relativistic electronic structure theory. It is capable of yielding highly accurate molecular integrals for all ranges of orbital parameters and quantum numbers.
\begin{description}
\item[Keywords]
Non-integer principal quantum numbers, Slater-type orbitals, Relativistic molecular auxiliary functions, Multi-center integrals
\item[PACS numbers]
... .
\end{description}
\end{abstract}
\maketitle

\section{\label{sec:intro}Introduction}

Formulae for interpreting visible (Balmer) and all electronic spectra of hydrogen were first caracterised by integers n$_1$ (visible n$_1$ =2) and n$_2$.
The idea of dropping the restriction on these integers was firstly suggested by Rydberg \cite{1_Rydberg_1980} in $1890$.
In the Bohr atom model (1913) these integers became known as 'quantum numbers' before full identification with the principle quantum number, with the solution of non$-$relativistic Schr{\"o}dinger equation for the Coulomb interaction in atomic units. This was much earlier than any attempt to develop a stable method for electronic structure calculation of many$-$electron systems or to construct a more flexible basis orbital to be used in this method. \\
The solution of non$-$relativistic Schr{\"o}dinger equation for the Coulomb interaction in atomic units $\left(a. u.\right)$, leads to expression for the wave$-$lengths $\lambda$ of spectral line emitted in a transition of the atom from quantum state $n_{2}$ to state $n_{1}$ \cite{2_Bethe_1957},
\begin{align}\label{eq:1_HLS}
\frac{1}{\lambda}=\frac{1}{2\pi}\left(E_{n_{2}}-E_{n_{1}} \right)=\frac{Z^2}{4\pi}
\left(\frac{1}{n_{1}^{2}} - \frac{1}{n_{2}^{2}}
\right),
\end{align}
where $Z$ is the nuclear charge, $E_{n_{1}}$, $E_{n_{2}}$ are the lower and upper energy levels for hydrogen$-$like atoms. $\left(n_{1}, n_{2} \right)$ are the principal quantum numbers. According to Bohr theory they have integer values. Rydberg through investigation of alkali spectra showed that for many$-$electron atoms empirically similar expression could be used:
\begin{align}\label{eq:2_MLS}
\frac{1}{\lambda}=\frac{1}{2\pi}\left(E_{n_{2}}-E_{n_{1}} \right)=\frac{Z^2}{4\pi}
\left(\frac{1}{\left(n_{1}-\delta_{1} \right)^{2}} - \frac{1}{\left(n_{2}-\delta_{2} \right)^{2}}
\right).
\end{align}
Here, $n^{*}=n-\delta$ is \textit{effective quantum number} with non$-$integer values. The quantity $\delta$ called \textit{the quantum defect} \cite{3_Seaton_1983}. It is depends on angular momentum quantum number $l$. The Eq. (\ref{eq:2_MLS}) was obtained by assumption that the $Z-1$ core electrons (which fill the inner shells of the core ion completely) do influence the energy of the one outermost electron by only screening the pure Coulomb potential. \cite{4_Hertel_2015, 5_Kostelecky_1985}. The analytical wave$-$function obtained from solution of Schr{\"o}dinger$-$like equation for the outermost electron with a Coulomb potential screened by $Z-1$ core electrons has a similar form as hydrogen atom solution but with non$-$integer values of principal quantum numbers \cite{5_Kostelecky_1985}. Note that, quantum mechanical justification of the Ritz's expansion \cite{6_Ritz_1903} for quantum defect finally, was given by Hartree \cite{7_Hartree_1928}. Such modification on Bohr's theory thus, guided the acquisition of the analytical Hartree$-$Fock SCF equations \cite{8_Roothaan_1951} and derivation of basis orbitals with non$-$integer principal quantum numbers as basis sets.

Many-electron atoms are well described by describing each electron by an 'atomic orbital' which corresponds physically to the hydrogen-like eigenfunction for that electron in the field of the 'bare' nucleus screened by all the other electrons. This can be visualised easily for such electronic configurations as the alkali metals, Li, Na, K etc, where the single electron in the highest 'n' shell evolves in an effective charge of approximately Z-1. In all cases, the idea is that a screening constant $\sigma$ can be ascribed to the other electrons on the basis of the 'average time spent' by this electron between the nucleus and the electron of interest. This approach is known as 'Slater screening cnstants' and the values tabulated give reasonable ionisation energies for the elements. It becomes necessary to modify the principan quantum number from n=4. However, the non-integer principal quantum number orbitals are hydrogen-like eigen-functions in this model, such that the energy of a given electron is given (in a.u.) by $-0.5*[(Z-\sigma)/n*]^2$ as it is for hydrogen and its eigenfunctions (here, $\sigma$ is the sum of shielding constants for all other electrons). In the 1930s, variational studies for the atoms with n=2 were carried out by Zener \cite{11_Zener_1930}. The hydrogen-like eigenfunctions are used but n (as well as the exponent) are variational parameters. The results show modest departure from n=2 and give a set of shielded exponents.

Slater in \cite{9_Slater_1930} realised that use of arbitrary basis functions is possible in the variational or Hartree method for atomic structure of many-electron systems. Later, the orbitals that carry his name were explicitly defined.

Slater$-$type orbitals with non$-$integer principal quantum numbers (NSTOs),
\begin{align}\label{eq:3_NSTO}
\chi\left(\zeta, \vec{r} \right)
=\frac{\left(2\zeta\right)^{n+1/2}}{\sqrt{\Gamma\left(2n+1\right)}}
r^{n-1}e^{-\zeta r}Y_{lm}\left(\theta, \varphi\right)
\end{align}
were originally considered as basis orbitals by Slater.
NSTOs are obtained by simplification of Laguerre functions in hydrogen-like wave$-$function (obtained from solution of Sch{\"o}dinger$-$like equation) by keeping only the term of the highest power of $r$. Compared to integer n STO, they provide extra flexibility for closer variational description of atoms and molecules since now in addition to orbital parameters $\zeta$ the principal quantum numbers are also variational parameters. This was already highlighted by Zener with a particular note that by putting extra parameters into the basis orbitals wherever flexibility may be obtained without increased complexity. Moreover, it was also predicted that using such sophisticated basis orbitals in molecular Hartree$-$Fock SCF calculations makes evaluation of the molecular integrals laborious. Absence of mathematical difficulties in evaluation of integrals for atomic systems on the other hand, allowed for research on using NSTOs \cite{12_Saturno_1958, 13_Synder_1960, 14_Allouche_1974} from different points of view. Complexity of basis sets was increased, effectiveness of such approximations on physical representation of a quantum mechanical system was investigated. It should be noted that work has been performed so far on electronic structure calculation of atoms and molecules with NSTOs, which are going to be investigated in detail below, clearly take inspiration from \cite{11_Zener_1930}. \\
Years after Zener's paper, the first attempt to calculate the molecules was carried out by Parr and Joy \cite{15_Parr_1957} in $1957$. In their immediately following work they revealed the bottleneck in solution of the integral evaluation problem that occurs in molecular calculations \cite{16_Joy_1958}. Four kinds of basic integral, the overlap, kinetic energy, nuclear attraction, electron$-$electron repulsion integrals were described. The prolate spheroidal coordinate method was used to evaluate these integrals from their expressions in terms of gamma functions, incomplete gamma functions, incomplete beta functions. The incomplete gamma functions on the other hand, have no explicit closed-form relations.

Following Parr, the single-center approach to molecular calculations using co-ordinate space was developed, in particular by Bishop \cite{35_Bishop_1963, 36_Bishop_1963, 37_Bishop_1965, 38_Bishop_1967}. This author is well-known for chemical applications of group theory and this seemingly guided his choice of molecule: $H_{3}O^{+}$, $CH_{4}$, $NH_{4}$ using Slater$-$type orbitals with non$-$integer principal quantum numbers. These studies suffer from a lack of mathematical and/or numerical tool-box for accurate calculation of molecular integrals over NSTOs.  Nevertheless, researchers had a desire to explore the advantages of using of NSTOs in molecular calculations.

Shortly afterwards, an alternative approach, based on momentum space was broached by Blanchard and Prosser in \cite{23_Prosser_1962} the convolution theorem for Fourier transform method was suggested for use alternatively to auxiliary functions methods (the method express the two$-$center integrals in ellipsoidal coordinates. The resulting simpler integrals are so called auxiliary functions). All subsequent studies so far in effort to derive an individual palatable technique for the solution, have been used either auxiliary functions method \cite{24_Saturno_1960, 25_Ludwig_1961, 26_Geller_1962} was originated by Hobson in \cite{27_Hobson_1931} and was suggested to use in quantum chemistry by Mulliken et al. in \cite{28_Mulliken_1949} or Fourier transform convolution method \cite{23_Prosser_1962} (this case results with highly oscillatory integrals involving spherical Bessel functions). As a sample rightfully, hydrogen molecule has been considered because it is the simplest possible molecule. It consists of two protons with either one$-$ or two$-$electrons. Mathematical elaboration the problem of integral evaluation within this framework was done by Geller and Silverstone in series of papers \cite{29_Geller_1963, 30_Geller_1963, 31_Geller_1964, 32_Geller_1964, 33_Silverstone_1966, 34_Silverstone_1967}.

 In another noteworthy attempt made by Allouche \cite{39_Allouche_1976} to eliminate the problem of integral evaluation taking into account the accuracy for results, was used the prolate spheroidal coordinates. The two$-$center one$-$electron integrals were again expressed in terms of auxiliary functions integrals involving the incomplete gamma functions. The two$-$center two$-$electron integrals were expressed in terms of auxiliary functions integrals involving the product of Legendre polynomials with different centers and the incomplete gamma functions. This time for integration of the resulting molecular auxiliary functions the numerical Gaussian quadrature procedure was suggested.
 Both the numerical procedure and its computer code makes it challenging to get accurate results for such auxiliary functions integrals given in \cite{39_Allouche_1976}. This may be because practical multi$-$precision libraries and the symbolic programming languages were not available when Allouche's paper was published. Even if they were available it would still be laborious (Please see \cite{40_Romanowski_2008, 41_Romanowski_2008, 42_Romanowski_2008, 43_Romanowski_2009} where numerical three$-$dimensional adaptive integration procedure used for calculation of two$-$center integrals with Slater$-$type functions (STOs). The principal quantum numbers restricted to be integers yet, even with lowest values of quantum numbers the results are insufficient). It should be noted that, a transformation method suggested for radial parts of NSTOs in \cite{39_Allouche_1976} as,
\begin{multline}\label{eq:4_NSTORT}
r_{B}^{n}e^{-\zeta r_{B}}\\
=\sqrt{2\pi}\sum_{l=0}^{\infty}\frac{1}{r_{A}R_{AB}}V_{n+2l}\left(r_{A},R_{AB},\zeta \right)Y_{l0}\left(\theta_{A}, \varphi_{A} \right),
\end{multline}
\begin{multline}\label{eq:5_NSTORTC}
V_{nl}\left(r_{A},R_{AB},\zeta\right)
=\int_{\vert R_{AB}-r_{A} \vert}^{R_{AB}+r_{A}}
r_{B}^{n}e^{-\zeta r}\mathscr{P}_{l0}\left(cos\hspace{0.3mm}\theta_{A} \right)dr_{B},
\end{multline}
shows that if the accuracy problem for two$-$ and three$-$center integrals are eliminated then, it is eliminated for four$-$center integrals as well. The last study before 90s that needs to be highlighted was performed by Taylor \cite{44_Taylor_1978}. There a general manipulation for inverse Gauss transforms were derived. Formulas for inverse Gauss transforms of Slater$-$type orbitals obtained in a previous research \cite{45_Kikuchi_1954} were generalized to NSTOs. As far as we know no detailed implementation of this method for molecular calculations yet. It is beyond scope of the present paper but, remains as an interesting work to be noted for future.

In 90s, the applications using NSTOs concentrated only on atomic implementations. It can be said that Koga and his co$-$workers played a dominant role in these applications \cite{46_Koga_1993, 47_Koga_1997, 48_Koga_1997, 49_Koga_1997, 50_Koga_1998, 51_Miguel_2000} through inclusively investigating the atoms for each individual modification on the basis sets. As stated above, the aim here was finding the optimal basis sets to be used in Hartree$-$Fock SCF calculations that represent the physical properties of the system as ideally as possible. For atoms with nuclear charge $Z$, $Z\leq 54$ detailed calculations using Slater$-$type basis orbitals had been performed by Clementi and Roetti in 1974 \cite{52_Clementi_1974}. In this study with single$-$, double$-$zeta basis set approximations the ground, excited states energies and linear combination coefficients of the atomic wave$-$function (in analytical solution of Hartree$-$Fock SCF equations the atomic wave$-$function is represented by linear combination of primitive basis orbitals) were perfected. Reference STO basis sets can be obtained from these studies, as well as from the numerical Slater-DFT code ADF that was developed by Baerends from 1973.

 Formally, studies on basis set construction methods \cite{53_Vega_1993, 54_Vega_1994, 55_Vega_1995, 56_Vega_1996, 57_Ema_1999} and re$-$optimization of orbital exponents \cite{58_Koga_1993, 59_Koga_1993, 60_Koga_1995} which are considered to improve the results given in \cite{52_Clementi_1974}, were performed using Slater$-$type orbitals. Thus, utilizing from the strategies developed in these works and dropping the restriction on principal quantum numbers provided further improvements. Note that, in early works of 90s it were believed that using NSTOs as basis orbitals in analytical solution of Hartree$-$Fock SCF equations may result in better values for energy then numerical solutions \cite{61_Bunge_1992}. In this class of computation, however, the main idea is testing the limits of used basis sets in terms of energy. The electron correlations and the relativistic effects are ignored. The best results for energy of atoms are found from numerical solution of Hartree$-$Fock SCF equations. Proximity of analytical solution to numerical solution thus, is so called Hartree$-$Fock limit of the used basis set approximation.\\
In following improvements for the results given in \cite{52_Clementi_1974}, inspiring from work in \cite{62_Hojer_1979} new variational parameters $\left(\eta \right)$ as $\left(r^{n-1}e^{-\zeta r^{\eta}} \right)$ \cite{63_Koga_1997} and from work in \cite{64_Szalewicz_1981} new functions $\left(cosh\left(\beta r + \gamma \right)\right)$ as $r^{n-1} e^{-\zeta r^{\eta}}cosh\left(\beta r +\gamma \right)$ \cite{65_Koga_1998, 66_Koga_1998, 67_Koga_1999} embedded to radial part of NSTOs. These new emendations led also to closer results to numerical solution for Hartree$-$Fock SCF. Besides, the quantity $\beta r$ in hyperbolic cosine was written as $\beta r^{\eta}$ \cite{68_Guseinov_2012}. Then it was decided to embedding the final generalized form of hyperbolic cosine functions as $cos_{pq}\left(\beta r^{\eta}+\gamma \right)$ \cite{69_Erturk_2015}. Using this kind of basis sets produced so the closest results to the numerical ones because each basis orbital in linear combination at least with four variational parameter and basis sets approximation such as double$-$zeta not yet is considered.

The improvements so far obtained for comprehensively investigating the physical properties of atoms via Hartree$-$Fock SCF method have been ended up with a decision about molecular applications of derived formulae and algorithms. A solution for the integrals evaluation problem in molecular calculations even with pure NSTOs without any additional parameter or functions is still pending. In the late of 90s one more attempt was made by Mekelleche and Baba$-$Ahmed \cite{70_Mekelleche_1997, 71_Mekelleche_2000}. Due to lack of benchmark values for the integrals then, a tremendous number of papers were published (Mostly by Guseinov, his co$-$workers \cite{72_Guseinov_2002, 73_Guseinov_2002, 74_Guseinov_2002, 75_Guseinov_2004, 76_Guseinov_2007, 77_Guseinov_2009, 78_Guseinov_2009} and Ozdogan, his co$-$workers \cite{79_Ozdogan_2002, 80_Ozdogan_2003, 81_Ozdogan_2003, 82_Ozdogan_2004}. We only cite here, those that are noteworthy. We refer the interested readers for more information \cite{83_Weniger_2007, 84_Weniger_2008, 85_Weniger_2012}). In almost all these works the ill$-$conditioned binomial series expansion,
\begin{align}\label{eq:6_MNBE}
\left(\xi \pm \nu \right)
=\sum_{s=0}^{\infty}\left(\pm 1\right)^{s}F_{s}\left(n\right)
\xi^{n-s} \nu^{s},
\end{align}
where, $F_{s}\left(n\right)$ are the binomial coefficients indexed by $n$, s is usually written
$\begin{pmatrix}
n\\
s
\end{pmatrix}$, with
\begin{align}\label{eq:7_BC}
\begin{pmatrix}
n\\
s
\end{pmatrix}
=\frac{\Gamma\left(n+1\right)}{\Gamma\left(s+1 \right)\Gamma\left(n-s+1\right)},
\end{align}
$\Gamma\left(n \right)$ is the gamma functions or the one$-$center expansion approximation,
\begin{align}\label{eq:8_NSTOOCE}
\chi_{nlm}\left(\zeta, \vec{r}\right)
=\lim_{N\rightarrow \infty}
\sum_{n'=l+1}^{N}V_{nl,n'l}^{N}\chi_{n'lm}\left(\zeta, \vec{r}\right)
\end{align}
was used. Here, $V_{nl,n'l}^{N}$ are the expansion coefficients used to represent the Slater$-$type orbitals with non$-$integer principal quantum numbers in terms of Slater$-$type orbitals with integer principal quantum numbers \cite{76_Guseinov_2007}. It was supposed that the results of calculations obtained with any of these approximations are accurate. Accordingly, these approximations have been applied for solution of various problems \cite{86_Guseinov_2002, 87_Guseinov_2003, 88_Guseinov_2004, 89_Ozdogan_2004, 90_Guseinov_2005, 91_Guseinov_2005, 92_Guseinov_2008, 93_Guseinov_2012, 94_Guseinov_2012}. \\
Finally, the benchmark results for two$-$, three$-$center one$-$ and two$-$electron molecular integrals which have been demanded for years in order to test the accuracy of any analytical method to be derived, obtained via numerical global-adaptive method through Gauss-Kronrod numerical integration extension \cite{95_Davis_1975} by us \cite{96_Bagci_2014, 97_Bagci_2015,98_Bagci_2015} with 35 correct decimals. A new molecular auxiliary functions introduced \cite{97_Bagci_2015}. Note that, the analytical evaluation of these functions involve some challenges namely, power functions with non-integer exponents, incomplete gamma functions and their multiplications have no explicit closed-form relations. The stability of results for the incomplete gamma functions varies according to domain of parameters \cite{17_Gil_2012} (please see also references therein). Efficient and accurate approximation for computation of the incomplete gamma functions still being studied in the literature \cite{18_Backeljauw_2014, 19_Nemes_2016, 20_Nemes_2015, 21_Nemes_2019, 22_Greengard_2019}. \\
In the following the definition and the origin of these new molecular auxiliary functions re-visited, an analytical method based on a recurrence strategy which is based on the criterion defined in our previous papers of the series, developed for computation of them \cite{99_Bagci_2018, 100_Bagci_2018}.

The main scope of the this paper is to show that neither in terms of accuracy nor CPU speed do disadvantages of using NSTOs in molecular calculations occur in comparison to Slater$-$type orbitals. The algorithm thereof, for the molecular auxiliary functions computation is detailed. A computer program code is written using $Julia$ programming language \cite{101_Bezanson_2017}. The results obtained for molecular auxiliary functions are compared with those obtained from numerical global-adaptive method based on Gauss-Kronrod numerical integration extension. The results obtained for the two$-$center integrals are compared with benchmark values given in \cite{96_Bagci_2014, 97_Bagci_2015}. Note that it would be a choice just presenting the formulae and sharing the details of computations upon request. The molecular integrals evaluation with NSTOs problem has preoccupied researchers for decades. As it is stated above plenty of research articles available in the literature that produce suspicious approaches and results. We thus, believe in this paper there should be sufficient information for both beginner and expert readers to re$-$compute the given formulas and re$-$produce the results. For some formulas, under$-$brace symbols are used in order to readily direct the readers to appendices where, the procedure for the computation of these formulas are specified. We finally hope this time the molecular integrals problem with non$-$integer Slater$-$type orbitals is reached an analytical solution.
\section{\label{sec:mol_aux} Revisiting the molecular auxiliary functions features}
The molecular auxiliary functions defined in prolate spheroidal coordinates ($\xi, \nu, \phi$) where, $1\leq\xi<\infty$, $-1\leq\nu\leq1$, $0\leq\phi\leq2\pi$, have the following form,
\begin{multline}\label{eq:9_RMA}
\left\lbrace \begin{array}{cc}
\mathcal{P}^{n_1,q}_{n_{2}n_{3}n_{4}}\left(p_{123} \right)
\\
\mathcal{Q}^{n_1,q}_{n_{2}n_{3}n_{4}}\left(p_{123} \right)
\end{array} \right\rbrace
\\
=\frac{p_{1}^{\sl n_{1}}}{\left({\sl n_{4}}-{\sl n_{1}} \right)_{\sl n_{1}}}
\int_{1}^{\infty}\int_{-1}^{1}{\left(\xi\nu \right)^{q}\left(\xi+\nu \right)^{\sl n_{2}}\left(\xi-\nu \right)^{\sl n_{3}}}\\ \times
\left\lbrace \begin{array}{cc}
P\left[{\sl n_{4}-n_{1}},p_{1}(\xi+\nu) \right]
\\
Q\left[{\sl n_{4}-n_{1}},p_{1}(\xi+\nu) \right]
\end{array} \right\rbrace
e^{p_{2}\xi-p_{3}\nu}d\xi d\nu,
\end{multline}
here, $\left\lbrace q, n_{1} \right\rbrace \in \mathbb{Z}$, $\left\lbrace n_{2}, n_{3}, n_{4}\right\rbrace \in \mathbb{R}$, $p_{123}=\left\lbrace p_{1}, p_{2}, p_{3}\right\rbrace$ (and in subsequent notation), $p_{1}>0$, $p_{2}>0$, $-p_{2}\leq p_{3} \leq p_{2}$.\\
$P,Q$ are the normalized complementary incomplete gamma and the normalized incomplete gamma functions\cite{102_Abramowitz_1972, 103_Temme_1994},
\begin{align}\label{eq:10_GPQ}
P\left[\alpha, z \right]
=\frac{\gamma\left(\alpha,z\right)}{\Gamma\left(\alpha\right)},
\hspace{5mm}
Q\left[\alpha, z \right]
=\frac{\Gamma\left(\alpha,z\right)}{\Gamma\left(\alpha\right)},
\end{align}
where $\gamma(a,z)$ and $\Gamma(a,z)$ are incomplete gamma functions,
\begin{align}\label{eq:11_GI}
\gamma\left(\alpha, z\right)
=\int_{0}^{z} t^{\alpha-1}e^{-t}dt,
\hspace{5mm}
\Gamma\left(\alpha, z\right)
=\int_{z}^{\infty} t^{\alpha-1}e^{-t}dt,
\end{align}
$\Gamma(a)$ is a complete gamma function,
\begin{align}\label{eq:12_GR}
\Gamma\left(\alpha\right)
=\Gamma\left(\alpha, z\right)+\gamma\left(\alpha, z\right),
\end{align}
and the Pochhammer's symbol $(\alpha)_{n}$ is defined,
\begin{equation}\label{eq:13_P}
\left(\alpha\right)_{n}
=\frac{\Gamma\left(\alpha+n\right)}{\Gamma\left(\alpha\right)}.
\end{equation}
The incomplete gamma functions in Eq. (\ref{eq:9_RMA}) arise as a result of two$-$electron interactions. As stated in our previous work, the symmetry properties of two-center two-electron integrals allow us to take advantage of the sum $P+Q=1$ instead of immediate expansion of incomplete gamma functions or using the relations $P=Q-1$, $Q=P-1$ with their conditional convergence \cite{103_Temme_1994}. This feature was formalized by a criterion given as,\vspace{1.5mm}\\
\textbf{Criterion.} Let $P\left[n_{4}-n_{1}, z \right]$ and $Q \left[n_{4}-n_{1}, z \right]$ then $n_{4}-n_{1}=a \pm c$, $n_{4}-n_{1}=a \pm d$, where $a \in \mathbb{R}$, $\left\lbrace c, d \right\rbrace \in \mathbb{Z}$ are true for any integrals that can be reduced to Eq. (\ref{eq:9_RMA}).\vspace{1.5mm}\\
It is now legitimate to use up$-$ and down$-$ward distant recurrence relations for normalized incomplete gamma functions \cite{103_Temme_1994} and reduce the Eq. (\ref{eq:9_RMA}) to well$-$known overlap$-$like integrals defined in prolate spheroidal coordinates \cite{100_Bagci_2018} which are independent from electron$-$electron interactions as follows,
\begin{multline}\label{eq:14_GQI}
\mathcal{G}^{n_{1},q}_{n_{2}n_{3}}\left(p_{123} \right)
=\frac{p_{1}^{n_{1}}}{\Gamma\left(n_{1}+1 \right)}
\\
\times \int_{1}^{\infty}\int_{-1}^{1} \left(\xi\nu\right)^{q}\left(\xi+\nu\right)^{n_{2}}\left(\xi-\nu \right)^{n_{3}}e^{-p_{2}\xi-p_{3}\nu}d\xi d\nu.
\end{multline}
By using the following relationship,
\begin{multline}\label{eq:15_QE}
\left(\xi \nu\right)^{q}\\
=\frac{1}{2^{2q}}\sum_{s_{1}}\left(-1\right)^{s_{1}}F_{s}\left(q\right)
\left(\xi+\nu \right)^{2q-2s_{1}} \left(\xi-\nu \right)^{2s_{1}},
\end{multline}
for Eq. (\ref{eq:14_GQI}) we have,
\begin{multline}\label{eq:16_GQE}
{\mathcal{G}^{n_{1},q}_{n_{2}n_{3}}}(p_{123})\\
=\frac{1}{2^{2q}}
\sum_{s_{1}}\left(-1\right)^{s_{1}}F_{s_{1}}\left(q \right)
{\mathcal{G}^{n_{1},0}_{n_{2}+2q-2s_{1}, n_{3}+2s_{1}}}(p_{123}),
\end{multline}
$\underline{0 \leq s_{1} \leq q}$. The $vectorization$ procedure which runs faster than the corresponding code containing loops, is used for computation of molecular auxiliary functions. Accordingly, it is more advantageous to detail the computer program code to be written in two sections. These sections are divided depending on the values of $p_{3}$ as $p_{3}=0$ and $p_{3} \neq 0$. Four sum indices are defined as follows,
\begin{itemize}
\item{$s_{1}$ is used in both sections three and four, for binomial expansion of $\left(\xi \nu\right)^{q}$}
\item{$s_{2}$ is used in section four, for series expansion of $e^{-p_{3}\nu}$}
\item{$s_{3}$ is used section four, for binomial expansion of $\left(\xi \nu\right)^{s_{2}}$}
\item{$s_{4}$ is used in both sections three and four, for series expansion of incomplete beta functions.}
\end{itemize}
These indices are results of explicitly writing the Eq. (\ref{eq:14_GQI}) by including all the sub$-$functions in its content. On the other hand, storing the value of all terms in explicit form of Eq. (\ref{eq:14_GQI}) requires using two indices (instead of four) run over a finite sum and an infinite sum, respectively. For this purpose we use an additional indices as $s_{5}$, $s_{6}$.
\section{\label{sec:mlc1} Computation of the molecular auxiliary functions. Case 1. {\normalsize $p_{3}=0$}.}
For $p_{3}=0$ we have the following relationship for the auxiliary functions given in the left hand sidde of Eq. ({\ref{eq:16_GQE}) with $q=0$ \cite{99_Bagci_2018},
\begin{multline}\label{eq:17_G0A}
\mathcal{G}^{n_{1},0}_{n_{2}n_{3}}\left(p_{120} \right)
=h^{n_{1},0}_{n_{2}n_{3}}\left(p_{12} \right)+ h^{n_{1},0}_{n_{3}n_{2}}\left(p_{12}\right)
\\
-k^{n_{1},0}_{n_{2}n_{3}}\left(p_{12}\right)
-k^{n_{1},0}_{n_{3}n_{2}}\left(p_{12}\right),
\end{multline}
here,
\begin{multline}\label{eq:18_G0h}
h^{n_{1},q'}_{n_{2}n_{3}}\left(p_{12}\right)
=\frac{p_{1}^{n_{1}}}{\Gamma\left(n_{1}+1 \right)}2^{n_{2}+n_{3}+q'+1}B\left(n_{2}+1,n_{3}+1\right)
\\
\times E_{-\left(n_{2}+n_{3}+q'+1\right)}\left(p_{2}\right)
-l^{n_{1},q'}_{n_{2}n_{3}}\left(p_{12}\right),
\end{multline}
\begin{multline}\label{eq:19_G0l}
l^{n_{1},q'}_{n_{2}n_{3}}\left(p_{12}\right)
=\frac{p_{1}^{n_{1}}}{\Gamma\left(n_{1}+1\right)} \\
\times \sum_{s_{4}}\frac{\left(-n_{2}\right)_{s_{4}}}{\left(n_{3}+s_{4}+1\right)s_{4}!}m^{n_{2}+q'-s_{4}}_{n_{3}+s_{4}+1}\left(p_{2}\right),
\end{multline}
where, $\underline{0 \leq s_{4} \leq N}$.
\begin{multline}\label{eq:20_G0m}
m^{n_{1}}_{n_{2}}\left(p\right)\\
=2^{n_{1}}U\left(n_{2}+1,n_{1}+n_{2}+2,p\right)\Gamma\left(n_{2}+1\right)e^{-p},
\end{multline}
and,
\begin{multline}\label{eq:21_G0k}
k^{n_{1},q'}_{n_{2},n_{3}}\left(p_{12}\right)
=\frac{p_{1}^{n_{1}}}{\Gamma\left(n_{1}+1 \right)}2^{n_{2}+n_{3}+q'+1} \\
\times B\left(n_{2}+1,n_{3}+1, \frac{1}{2}\right)
E_{-\left(n_{2}+n_{3}+q'+1\right)}\left(p_{2}\right).
\end{multline}
with, confluent hypergeometric functions of first kind \cite{102_Abramowitz_1972} and $B\left(\alpha,\beta, z\right)$ incomplete beta functions,
\begin{align}\label{eq:22_BI}
B\left(\alpha,\beta, z\right)
=\int_{0}^{z}t^{\alpha-1}(1-t)^{\beta-1}dt.
\end{align}
Note that, while $q=0$, $q'=0$. The Eqs. (\ref{eq:18_G0h}-\ref{eq:21_G0k}) are given in general form and be used for $q \neq 0$. The $q'$ in fact, is a sum indices that arises from series expansion of exponential functions. We discuss this in the following section but such generalization of the Eqs. (\ref{eq:18_G0h}-\ref{eq:21_G0k}) is to avoid duplication.\\
The simplified form of Eq. (\ref{eq:17_G0A}) that is easier to use in coding, is written as follows,
\begin{multline}\label{eq:23_G0AS}
\mathcal{G}^{n_{1},0}_{n_{2}n_{3}}\left(p_{120} \right)
=\frac{p_{1}^{n_{1}}}{\Gamma\left(n_{1}+1 \right)}
2^{n_{2}+n_{3}+1}
\underbrace{B_{n_{2}+1,n_{3}+1}}_{\mathcal{G}_{01}} \\
\times \underbrace{E_{-\left(n_{2}+n_{3}+1 \right)}\left(p_{2}\right)}_{\mathcal{G}_{02}}
-\underbrace{l_{n_{2}n_{3}}^{n_{1},0}\left(p_{12}\right)}_{\mathcal{G}_{03}}
-\underbrace{l_{n_{3}n_{2}}^{n_{1},0}\left(p_{12}\right)}_{\mathcal{G}_{04}}.
\end{multline}
Considering the Eq. (\ref{eq:16_GQE}), the values for $n_{1}$ $n_{2}$, $n_{3}$ in the Eq. (\ref{eq:23_G0AS}) should replace with $n_{1} \rightarrow n_{1}$, $n_{2} \rightarrow n_{2}+2q-2s_{1}$, $n_{3} \rightarrow n_{3}+2s_{1}$. The second under$-$braced function is called to as $\mathcal{G}_{02}$, represents the generalized exponential integral functions \cite{102_Abramowitz_1972}. It is invariant for each term of the summation. \\
The following recurrence relationship is derived for the beta function depicted with $\mathcal{G}^{01}$
\begin{multline}\label{eq:24_BR}
B_{z-2s,z'+2s}
=\frac{\left(z'+2s-1 \right)\left(z'+2s-2 \right)}{\left(z-2s \right)\left(z-2s+1 \right)} \\
\times B_{z-2s+2,z'+2s-2},
\end{multline}
where, $z=n_{2}+2q+1$, $z'=n_{3}+1$.\\
The $\mathcal{G}_{03}$ is a re$-$written form of Eq. (\ref{eq:19_G0l}) according to Eq. (\ref{eq:16_GQE}):
\begin{multline}\label{eq:25_G0lE}
l^{n_{1},0}_{n_{2}+2q-2s_{1}, n_{3}+2s_{1}}\left(p_{12}\right)
=\frac{p_{1}^{n_{1}}}{\Gamma\left(n_{1}+1\right)} \\
\times \sum_{s_{4}=0}^{\infty}
\underbrace{\frac{\left[-\left(n_{2}+2q-2s_{1}\right)\right]_{s_{4}}}
{\left(n_{3}+2s_{1}+s_{4}+1\right)}}_{P_{1}}
\underbrace{
\frac{m^{n_{2}+2q-2s_{1}-s_{4}}_{n_{3}+2s_{1}+s_{4}+1}\left(p_{2}\right)}{s_{4}!}}_{M_{1}}.
\end{multline}
The expression for $\mathcal{G}^{04}$ is obtained only by exchanging the indices $n_{2}+2q-2s_{1}$ and $n_{3}+2s_{1}$ so the under$-$brace symbols in this case become $P_{2}$, $M_{2}$.
The following form of down$-$, up$-$ward recurrence relationships then, are used for Pochhammer's symbols in $P_{1}$ and $P_{2}$,
\begin{multline}\label{eq:26_PR1}
\left(z-2s\right)_{s'}
=\frac{\left(z-2s \right) \left(z-2s+1 \right)}
{\left(z-2s+s' \right)\left(z-2s+s'+1 \right)} \\
\times \left(z-2s+2 \right)_{s'},
\end{multline}
\begin{multline}\label{eq:27_PR2}
\left(z+2s\right)_{s'}
=\frac{\left(z+2s+s'-1 \right) \left(z+2s+s'-2 \right)}
{\left(z+2s-1 \right)\left(z+2s-2 \right)} \\
\times \left(z+2s-2 \right)_{s'}.
\end{multline}
The $m$ functions with two sum indices are reduced to one and computed with using recurrence relationships:
\begin{multline}\label{eq:28_mR}
\underbrace{m^{n_{2}+2q-s}_{n_{3}+s+1}\left(p_{2} \right)}_{m^{1}, n_{2}\rightarrow n_{3}\Rightarrow m^{2}}
=\underbrace{\frac{1}{4}\frac{\left(n_{3}+s \right)}{\left( n_{3}+2q-s+1 \right)}}_{c^{11}, n_{2}\rightarrow n_{3}\Rightarrow c^{21}}
\underbrace{m^{n_{2}+2q-\left(s-2 \right)}_{n_{3}+s-1}\left(p_{2}\right)}_{m^{11}, n_{2}\rightarrow n_{3}\Rightarrow m^{21}}
\\
+\underbrace{\frac{1}{2}\frac{\left(n_{2}-n_{3}+2q-2s-p_{2}+1 \right)}{\left(n_{2}+2q-s+1\right)}}_{c^{21}, n_{2}\rightarrow n_{3}\Rightarrow c^{22}}
\underbrace{m^{n_{2}+2q-\left(s-1\right)}_{n_{3}+s}\left(p_{2} \right)}_{m^{21}, n_{2}\rightarrow n_{3}\Rightarrow m^{22}}.
\end{multline}
In the Eq. (\ref{eq:25_G0lE}) for $M_{2}$, $n_{2}$ are exchanged with $n_{3}$.\\
The readers should look to Appendix \ref{appb:Vec_form_mlc1} for vectorized forms of the equations presented in this section.
\section{\label{sec:mlc2} Computation of the molecular auxiliary functions. Case 2. {\normalsize $p_{3} \neq 0$}.}
The derived relationships in our previous papers \cite{99_Bagci_2018, 100_Bagci_2018} for molecular auxiliary functions while $p_{3} \neq 0$ can be recapitulated in the present paper. The Eq. (\ref{eq:16_GQE}) is still primary but its reduced form $\left(\mathcal{G}^{n_{1},0}\right)$ on the right$-$hand side, is as follows,
\begin{align}\label{eq:29_GQJ}
\mathcal{G}^{n_{1},0}_{n_{2},n_{3}}\left(p_{123} \right)
=\frac{p_{1}^{n_{1}}}{\Gamma\left(n_{1}+1 \right)}
\sum_{s} \left(-1\right)^{s}
\mathcal{J}_{n_{2},n_{3}}^{s,s,s}\left(p_{32} \right),
\end{align}
\begin{multline}\label{eq:30_JSSS}
\mathcal{J}_{n_{2},n_{3}}^{s,s,s}\left(p_{32} \right) \\
=\frac{1}{2^{s}}\sum_{s'} \left(-1\right)^{s'}F_{s'}\left(s \right)
\mathcal{J}_{n_{2}+2s-2s',n_{3}+2s'}^{s,s,0}\left(p_{32} \right),
\end{multline}
\begin{multline}\label{eq:31_JSS0}
\mathcal{J}^{s,s,0}_{n_{2}n_{3}}\left(p_{12} \right)
=\frac{p_{1}^{n_{1}}}{\Gamma\left(n_{1}+1 \right)}
2^{n_{2}+n_{3}-s+1}
\underbrace{B_{n_{2}+1,n_{3}+1}}_{\mathcal{J}_{01}} \\
\times \underbrace{E_{-\left(n_{2}+n_{3}-s+1 \right)}}_{\mathcal{J}_{02}}
-\underbrace{l_{n_{2}n_{3}}^{n_{1},-s}\left(p_{12}\right)}_{\mathcal{J}_{03}}
-\underbrace{l_{n_{3}n_{2}}^{n_{1},-s}\left(p_{12}\right)}_{\mathcal{J}_{04}},
\end{multline}
Here, $s=s_{2}$, $n_{2}=n_{2}+2q-2s_{1}+2s_{2}-2s_{3}$, $n_{3}=n_{3}+2s_{1}+2s_{3}$. Before starting the discussion on the computational procedure, it should be noticed that same as previous section, the Eq. (\ref{eq:16_GQE}) is computed by directly inserting the Eqs. (\ref{eq:30_JSSS}, \ref{eq:31_JSS0}) into it. Instead of producing multiple functions that need to be computed, this is recommended in our study based on vectorization procedure.\\
Starting with $\mathcal{J}_{01}$ of the Eq. (\ref{eq:31_JSS0}) the following recurrence relationships are used first,
\begin{multline}\label{eq:32_JBR1}
B_{z+2s, z'}\\
=
\underbrace{\frac{\left(z+2s-1\right) \left(z+2s-2\right)}
{\left(z+z'+2s-1\right)\left(z+z'+2s-2\right)}
}_{br}
B_{z+2s-2,z'},
\end{multline}
with $z=n_{2}+2q-s$, $z=n_{3}+s$, and,
\begin{multline}\label{eq:33_JBR2}
B_{z-s, z'+s}\\
=
\underbrace{
\frac{\left(z'+2s-1\right) \left(z'+2s-2\right)}
{\left(z-2s\right)\left(z-2s+1\right)}
}_{bc}
B_{z-s+2,z'+s-2},
\end{multline}
with, $z=n_{2}+2q+2s$, $z'=n_{3}$.\\
The $\mathcal{J}_{02}$ represents the exponential integrals functions. They are computed as follows,
\begin{align}\label{eq:34_JER}
E_{-a-s}\left(p\right)
=\frac{1}{p}
\left\lbrace
e^{-p} + \left(a+s \right)E_{-a-s+1}\left(p\right)
\right\rbrace,
\end{align}
here, $-\left(a+s\right)=-\left(n_{2}+2q+n_{3}\right)-s_{2}$, $p=p_{2}$.\\
The auxiliary functions given with $\mathcal{J}_{03}$ and $\mathcal{J}_{04}$ are the most challenging to compute since they are defined with four sum indices. They have symmetry that allows us to represent them with only two sum indices (one use for finite sum and the other for infinite sum). Note that, the molecular auxiliary functions derived for evaluation of integrals over Slater$-$type orbitals with integer principal quantum number have two sum indices of the same property. This is the most important feature of our method. The main reason for the claim that for both accuracy and CPU speed there should be no disadvantages of using NSTOs in molecular calculations. Taking into account a small computationally meaningful modification on the Eq. (\ref{eq:19_G0l}) the $\mathcal{J}_{03}$ and $\mathcal{J}_{04}$ are expressed as,
\begin{multline}\label{eq:35_JlE}
l^{n_{1},q}_{n_{2},n_{3}}\left(p_{12} \right)
=\frac{p_{1}^{n_{1}}}{\Gamma\left(n_{1}+1 \right)}
\\
\times \overbrace{\left(-n_{2} \right)_{-q}}^{pe_{1}}
\sum_{s}
\frac{\overbrace{\left(-n_{2}-q \right)_{s}}^{pl_{1}}}
{\left(n_{3}+s+1\right)s!}
\frac{\overbrace{m^{\left(n_{2}+q \right)}_{n_{3}+s+1}\left(p_{2} \right)}^{m_{1}}}
{\underbrace{ \left(-n_{2}+s\right)_{-q}}_{ps_{1}}}
\end{multline}
where, $s=s_{4}$, $n_{1}=s_{2}$, $q=-s_{2}$, $p_{12}=p_{32}$, $p_{1}=p_{3}$. For $\mathcal{J}_{03}$ function, $n_{2}=n_{2}+2q-2s_{1}+2s_{2}-2s_{3}$, $n_{3}=n_{3}+2s_{1}+2s_{3}$. The values of $n_{2}$ with $n_{3}$ and the symbols $m_{1}$, $pe_{1}$, $pl_{1}$, $ps_{1}$ with $m_{2}$, $pe_{2}$, $pl_{2}$, $ps_{2}$ are exchanged respectively for $\mathcal{J}_{04}$. \\
The recurrence relationship to be used for computation of $ps^{1}$, $pe^{1}$ determined as,
\begin{multline}\label{eq:36_JPR}
\left(-\left[n_{2}+2q\right]-2s\right)_{s}
\\
=\underbrace{
\frac{\left[-\left(n_{2}+2q\right)-2s\right]}
{\left[-\left(n_{2}+2q\right)-s\right]}
\left[-\left(n_{2}+2q\right)-2s+1\right]
}_{psr_{1}, n_{2}\rightarrow n_{3} \Rightarrow psr_{2}}
\\
\left(-\left[n_{2}+2q\right]-2s+2\right)_{s-1},
\end{multline}
\begin{multline}\label{eq:37_JPR}
\left(-\left[n_{2}+2q\right]-2s+s'\right)_{s'}
\\
=\underbrace{\frac{\left[-\left(n_{2}+2q\right)-2s+s'+s-1\right]}
{\left[-\left(n_{2}+2q\right)-2s+s'-1\right]}}_{psc_{1}, n_{2}\rightarrow n_{3} \Rightarrow psc_{2}}
\\
\times \left(-\left[n_{2}+2q\right]-2s+s'-1\right)_{s'}.
\end{multline}
There is no need for an additional effort to compute $pe_{1}$, $pe_{2}$ since $ps_{1}, ps_{2}$ contains all of their terms. $pl_{1}$ and $pl_{2} \left(n_{2}\rightarrow n_{3}\right)$ are in same form with $p_{m}$, thus same procedure for computation of them are used. This also applies for factorial and binomial coefficients. The only functions remain that require special attention, are $m_{1}$, $m_{2}$. The sum indices reduced recurrence relationships expressions for $m$ functions are as follows,
\begin{multline}\label{eq:38_Jms}
\underbrace{m^{\left(n_{2}+s\right)+2q-s'}_{n_{3}+s'+1}\left(p \right)}_{m_{1}, n_{2}\rightarrow n_{3}\Rightarrow m_{2}}
=\frac{1}{4}\underbrace{\frac{\left(n_{3}+s' \right)}{\left(\left(n_{3}+s\right)+2q-s'+1 \right]}}_{c_{11}, n_{2}\rightarrow n_{3}\Rightarrow c_{21}}
\\
\times \underbrace{m^{\left(n_{2}+s\right)+2q-\left(s'-2 \right)}_{n_{3}+s'+1}\left(p\right)}_{m_{11}, n_{2}\rightarrow n_{3}\Rightarrow m_{21}}
\\
+\frac{1}{2}\underbrace{\frac{\left[\left(n_{2}+s\right)-n_{3}+2q-2s'-p_{2}+1 \right]}{\left(\left(n_{2}+s\right)+2q-s'+1\right]}}_{c_{21}, n_{2}\rightarrow n_{3}\Rightarrow c_{22}}
\\
\times \underbrace{m^{\left(n_{2}+s\right)+2q-\left(s'-1\right)}_{n_{3}+s'}\left(p \right)}_{m_{21}, n_{2}\rightarrow n_{3}\Rightarrow m_{22}},
\end{multline}
\begin{multline}\label{eq:39_Jmsu}
\underbrace{m^{\left(n_{2}+s\right)+2q-s'}_{n_{3}+s'+1}\left(p \right)}_{m_{1}, n_{2}\rightarrow n_{3}\Rightarrow m_{2}}
=4\underbrace{\frac{\left(n_{2}+s\right)+n_{3}+2q+p+1}{p}}_{r_{11}, n_{2}\rightarrow n_{3}\Rightarrow r_{21}}
\\
\times \underbrace{m^{\left(n_{2}+s\right)+2q-\left(s'+1 \right)}_{n_{3}+s'+1}\left(p\right)}_{m_{11}, n_{2}\rightarrow n_{3}\Rightarrow m_{21}}
\\
+2\underbrace{\frac{s'-\left(n_{2}+s\right)-2q+1}{p}}_{r^{21}, n_{2}\rightarrow n_{3}\Rightarrow r_{22}}
\\
\times \underbrace{m^{\left(n_{2}+s\right)+2q-\left(s'+2\right)}_{n_{3}+s'+1}\left(p \right)}_{m_{21}, n_{2}\rightarrow n_{3}\Rightarrow m_{22}}.
\end{multline}
All the terms stored in the memory are re-collected in line with the Eqs. (\ref{eq:31_JSS0}, \ref{eq:35_JlE}) then they are used in explicit form of the Eq. (\ref{eq:16_GQE}) given below,
\begin{multline}\label{eq:40_GQC}
G^{n_{1},q}_{n_{2},n_{3}}\left(p_{123} \right)
=\frac{1}{2^{q}}\sum_{s_{1},s_{2},s_{3}} \left(-1\right)^{s_{1}+s_{4}+s_{2}}F_{s_{1}}\left(q\right)
\frac{1}{2^{s_{4}}}
\\
\times F_{s_{3}}\left(s_{2}\right)
\left\lbrace
\frac{1}{2^{-s_{2}}}
\left(
\frac{p_{3}^{s_{2}}}{\Gamma\left(s_{2}+1\right)}
2^{n_{2}+n_{3}+2q+s_{2}+1}
\right.
\right.
\\
\times B_{n_{2}+2q-2s_{1}+2s_{2}-2s_{3}+1, n_{3}+2s_{1}+2s_{3}+1}
\\
\times E_{-\left(n_{2}+n_{3}+2q+s_{2}+1 \right)}\left(p_{2}\right)
\\
-l^{s_{2}-s_{2}}_{n_{2}+2q-2s_{1}+2s_{2}-2s_{3}, n_{3}+2s_{1}+2s_{2}}\left(p_{32}\right)
\\
-l^{s_{2}-s_{2}}_{n_{3}+2s_{1}+2s_{2}, n_{2}+2q-2s_{1}+2s_{2}-2s_{3}}\left(p_{32}\right)
\left.
\left.
\right)
\right\rbrace.
\end{multline}
The domain of $s_{6}$ sum index is dependent to upper limit of summation which must covers all the terms in Eq. (\ref{eq:40_GQC}).\\
For vectorized forms of the equations presented in this section see Appendix \ref{appc:Vec_form_mlc2}.
\section{\label{sec:res_discuss} Results and Discussions}
\begin{table*}[htp!]
\caption{\label{tab:OverlapTC} The comparative values for the two$-$center overlap integrals over non$-$integer Slater$-$type orbitals.}
\begin{ruledtabular}
\begin{tabular}{ccccccccc}
$n$ & $l$ & $n'$ & $l'$ & $\lambda$ & $\rho$ & $\tau$ & Results
\\
\hline
$50.1$ & $0$ & $50.0$ & $0$ & $0$ & $5.1$ & $0$ & \begin{tabular}[c]{@{}l@{}l@{}l@{}l@{}l@{}l@{}l@{}}
\hspace{1mm}9.57914 65146 38189 77903 14416 92566 55702 E$-$01(35)\footnotemark[1]\\
\hspace{1mm}9.57914 65146 38189 77903 14416 92566 55702 E$-$01(80)\footnotemark[2]\\
\end{tabular}
\vspace{1mm}
\\
$50.1$ & $1$ & $50.0$ & $1$ & $1$ & $5.1$ & $0$ & \begin{tabular}[c]{@{}l@{}l@{}l@{}l@{}l@{}l@{}l@{}}
\hspace{1mm}9.72384 17544 16182 68349 60818 96583 16551 E$-$01(35)\footnotemark[1]\\
\hspace{1mm}9.72384 17544 16182 68349 60818 96583 16551 E$-$01(80)\footnotemark[2]\\
\end{tabular}
\vspace{1mm}
\\
$50.1$ & $2$ & $50.0$ & $2$ & $0$ & $5.1$ & $0$ & \begin{tabular}[c]{@{}l@{}l@{}l@{}l@{}l@{}l@{}l@{}}
\hspace{1mm}9.18991 91933 69431 71198 54394 33099 01583 E$-$01(35)\footnotemark[1]\\
\hspace{1mm}9.18991 91933 69431 71198 54394 33099 01583 E$-$01(80)\footnotemark[2]\\
\end{tabular}
\vspace{1mm}
\\
$50.1$ & $3$ & $50.0$ & $3$ & $1$ & $5.1$ & $0$ & \begin{tabular}[c]{@{}l@{}l@{}l@{}l@{}l@{}l@{}l@{}}
\hspace{1mm}9.13139 56806 63425 61199 88426 62258 19702 E$-$01(35)\footnotemark[1]\\
\hspace{1mm}9.13139 56806 63425 61199 88426 62258 19703 E$-$01(80)\footnotemark[2]\\
\end{tabular}
\vspace{1mm}
\\
$50.1$ & $4$ & $50.0$ & $4$ & $2$ & $5.1$ & $0$ & \begin{tabular}[c]{@{}l@{}l@{}l@{}l@{}l@{}l@{}l@{}}
\hspace{1mm}9.08402 38184 02459 97117 43224 82911 00064 E$-$01(40)\footnotemark[1]\\
\hspace{1mm}9.08402 38184 02459 97117 43224 82911 00064 E$-$01(85)\footnotemark[2]\\
\end{tabular}
\vspace{1mm}
\\
$50.1$ & $5$ & $50.0$ & $5$ & $1$ & $5.1$ & $0$ & \begin{tabular}[c]{@{}l@{}l@{}l@{}l@{}l@{}l@{}l@{}}
\hspace{1mm}8.66195 74959 32514 08801 94794 08243 70292 E$-$01(40)\footnotemark[1]\\
\hspace{1mm}8.66195 74959 32514 08801 94794 08243 70292 E$-$01(85)\footnotemark[2]\\
\end{tabular}
\footnotetext[0]{The numbers in pharantesis represent the upper limit of summation $N$.}
\footnotetext[1]{Benchmark result obtained via global-adaptive method with Gauss-Kronrod extension.}
\footnotetext[2]{Results obtained via Eq. (\ref{eq:40_GQC}).}			
\end{tabular}
\end{ruledtabular}
\end{table*}

\begin{table*}[htp!]
\caption{\label{tab:OverlapTCC} Convergence behavior of the analytical solution for two$-$center overlap integrals.}
\begin{ruledtabular}
\begin{tabular}{ccccccccc}
$n$ & $l$ & $n'$ & $l'$ & $\lambda$ & $\rho$ & $\tau$ & Results
\\
\hline	
$50.1$ & $5$ & $50.0$ & $5$ & $1$ & $5.1$ & $0$ &
\begin{tabular}[c]{@{}l@{}l@{}l@{}l@{}l@{}l@{}l@{}}		
\hspace{1mm}1.32568 13525 90586 81329 51435 96242 73791 E$+$36(10)\\
-2.90971 87630 90363 23342 45215 08552 63018 E$+$34(15)\\
\hspace{1mm}1.23299 50039 89889 33472 47029 50376 17086 E$+$32(20)\\
-8.07987 96650 25341 37309 26554 15979 78721 E$+$28(25)\\
\hspace{1mm}1.93904 45950 34950 77981 80976 64495 06345 E$+$24(30)\\
\hspace{1mm}1.64177 83542 71254 80306 89257 94945 30469 E$+$20(35)\\
\hspace{1mm}2.24224 67720 72493 39894 23507 37260 10690 E$+$15(40)\\
\hspace{1mm}8.10564 74767 07974 06257 20626 75989 83928 E$+$08(45)\\
-1.56878 26202 66460 85553 98231 88490 25729 E$+$01(50)\\
\hspace{1mm}8.66195 75005 97684 59795 29121 58447 25427 E$-$01(55)\\
\hspace{1mm}8.66195 74959 32514 00384 71911 22855 86889 E$-$01(60)\\
\hspace{1mm}8.66195 74959 32514 08801 68929 31385 75004 E$-$01(65)\\
\hspace{1mm}8.66195 74959 32514 08801 94793 64570 45165 E$-$01(70)\\
\hspace{1mm}8.66195 74959 32514 08801 94794 08242 21380 E$-$01(75)\\
\hspace{1mm}8.66195 74959 32514 08801 94794 08243 70283 E$-$01(80)\\
\hspace{1mm}8.66195 74959 32514 08801 94794 08243 70292 E$-$01(85)\\
\hspace{1mm}8.66195 74959 32514 08801 94794 08243 70292 E$-$01(90)\\	
\end{tabular}
\end{tabular}
\footnotetext[0]{The numbers in pharantesis represent the upper limit of summation $N$.}
\end{ruledtabular}
\end{table*}
An efficient method for computation of the relativistic molecular auxiliary functions given in the Eq. (\ref{eq:9_RMA}) is presented. They are reduced to Eq. (\ref{eq:17_G0A}) and Eq. (\ref{eq:29_GQJ}) according to the criterion that represent the symmetry of two$-$electron interactions. This is also consistent with the idea that the overlap integrals are basic building block for molecular integrals since the Eqs. (\ref{eq:17_G0A}, \ref{eq:29_GQJ}) are in fact the representation of two$-$center overlap integrals in prolate spheroidal coordinates. This simply provide the necessary and sufficient condition to prove the accuracy of proposed fully analytical method. The two$-$center overlap integrals of non$-$integer Slater$-$type orbitals in the lined$-$up coordinate systems are given as,
\begin{align}\label{eq:41_Overlap}
		S_{nl\lambda,n'l'\lambda}(\rho, \tau)
		=\int
		\chi^{*}_{nl\lambda}\left(\zeta, \vec{r}_{A}\right)
		\chi_{n'l'\lambda}\left(\zeta', \vec{r}_{B}\right)dV
\end{align}
with,
\begin{align*}
	\rho = \frac{R}{2}\left(\zeta+\zeta' \right),
	\hspace*{1mm}
	\tau = \frac{\zeta-\zeta'}{\zeta+\zeta'}
\end{align*}
and, $\vec{R}=\vec{R}_{AB}=\vec{r}_{A}-\vec{r}_{B}$, $\zeta, \zeta'$ are orbital parameters. Please see \cite{96_Bagci_2014} (and references therein) for explicit form of Eq. (\ref{eq:41_Overlap}). For calculation of the Eq. (\ref{eq:41_Overlap}) the following form accordingly, is used,
\begin{multline}\label{eq:42_Overlap}
	S_{nl\lambda,n'l'\lambda}(\rho, \tau)=
	N_{nn'}\left(\rho, \tau \right)\sum_{a}^{l}\sum_{b=\lambda}^{l'}\sum_{c=0}^{a+b}
	g_{ab}^{c}\left(l\lambda,l'\lambda \right)
	\\
	\times \mathcal{P}_{n-a,n'-b,0}^{0,c}\left(0,\rho, \tau \right).
\end{multline}
The $g^{c}_{ab}$ coefficients arise from product of two spherical harmonics with different centers \cite{72_Guseinov_2002}. The results of calculations presented in Tables \ref{tab:OverlapTC} and \ref{tab:OverlapTCC} are obtained from the Eq. (\ref{eq:42_Overlap}). It is clear from this equation that the accuracy of the used method should be tested by increasing the values of angular momentum coefficients $l,l'$ and considering as high as possible values for principal quantum numbers $n,n'$. In the tables presented in this study, the fixed values for principal quantum numbers, $n=50.1$, $n'=50.0$ are used. The values of orbital parameters are chosen as $\rho=5.1$, $\tau=0$. These values are based on experience from previous calculations which lead to know that methods hitherto developed in order to calculate the two$-$center overlap integrals with non$-$integer principal quantum numbers have failed. The values of angular momentum quantum numbers $l,l'$ are increased from $\left\lbrace l,l' \right\rbrace=0$ to $\left\lbrace l,l' \right\rbrace=5$, respectively. Benchmark results are obtained via numerical global$-$adaptive method with Gauss$-$Kronrod numerical integration extension using Mathematica programming language \cite{104_MathematicaProg}. The calculations are performed with 40$-$digit accuracy by setting the \textbf{WorkingPrecision} option to $40$. This causes all internal computations to be done to 40$-$digit precision. \\
The $Julia$ programming language \cite{105_Bezanson_2017,106_Bagci_2020} is used for computation of the fully analytical method. This programming language allows easy use of this existing code written in $C$ or $Fortran$ programming languages. It has a "no boilerplate" philosophy: functions can be called directly from it without any "glue" code, code generation, or compilation even from the interactive prompt. This is accomplished by making an appropriate call with $ccall$, which looks like an ordinary function call. The most common syntax for $ccall$ is as follow,
\begin{multline*}
	ccall(
	(symbol,library),\\
	RetType,
	(ArgType1, ...),
	Arg1, ...
	).
\end{multline*}
For accuracy only an additional computer algebra package, called {\sl Nemo} \cite{107_Fieker_2017} is required. This package is based on $C$ libraries such as $FLINT, ANTIC, Arb, Pari$ and $Singular$. It has a module system which is use to provide access to $Nemo$. It is imported and used all exported functionality by simply type $using$ $Nemo$.\\
In the light of the previous sections, for fully analytical method the upper limit of summation $N$ that gives results equivalent to numerical ones has found to be $N=85$.

\appendix
\section{\label{appb:Vec_form_mlc1} Vectorized form for the functions in Section \ref{sec:mlc1}}
The beta functions are now be written in vector form as,
\begin{align}\label{eq:a1_BRV}
	\boxed{B[s_{1}+1]
		=\frac{\left(z'+2s_{1}-1 \right)\left(z'+2s_{1}-2 \right)}{\left(z-2s_{1} \right)\left(z-2s_{1}+1 \right)}
		B[s_{1}]}
\end{align}
In order to compute the Pochhammer's symbols in $P_{1}$, $P_{2}$, first, $s_{1}=0$ is considered.
\begin{align}\label{eq:a2_P1P2p1p2V}
	\boxed{
		\begin{array}{c}
			P_{1}[s_{4},1]=p_{1}[s_{4}]/ \left(n_{3}+s_{4}+1 \right)
			\vspace{1mm} \\
			P_{2}[s_{4},1]=p_{2}[s_{4}]/ \left(n_{2}+2q+s_{4}+1 \right)
		\end{array}
	}
\end{align}
where,
\begin{align}\label{eq:a3_p1p2V}
	\begin{array}{c}
		p_{1}[s_{4}+1]=-\left(n_{2}+2q-s_{4} + 1 \right)p_{1}[s_{4}]
		\vspace{1mm} \\
		p_{2}[s_{4}+1]=-\left(n_{3}-s_{4}+ 1\right)p_{2}[s_{4}]
	\end{array}
\end{align}
In a vector form considering the Eqs. (\ref{eq:25_G0lE}$-$\ref{eq:27_PR2}), They are given as,
\begin{multline}\label{eq:a4_P1V}
	\boxed{
		P_{1}\left[s_{4},s_{1}+1\right]}
	\\
	\boxed{
		=\frac{\left(z+2s_{1}+s_{4}-1 \right) \left(z+2s_{1}+s_{4}-2 \right)}
		{\left(z+2s_{1} \right)\left(z+2s_{1}-1 \right)}
		P_{1}[s_{4},s_{1}]
	}
\end{multline}
\begin{multline}\label{eq:a5_P2V}
	\boxed{
		P_{2}\left[s_{4}, s_{1}+1\right]}
	\\
	\boxed{
		=\frac{\left(z'-2s_{1}-1 \right) \left(z'-2s_{1} \right)}
		{\left(z'-2s_{1}+s_{4}-2 \right)\left(z'-2s_{1}+s_{4}-1 \right)}
		P_{2}[s_{4},s_{1}]
	}
\end{multline}
\begin{align}\label{eq:a6_m1m2M1M2V}
	\boxed{
		\begin{array}{c}
			m_{1}[s_{5}+1]=\frac{1}{4}c^{11}m_{11}[s_{5}]+\frac{1}{2}c^{12}m_{12}[s_{5}]
			\vspace{1mm} \\
			m_{2}[s_{5}+1]=\frac{1}{4}c^{21}m_{21}[s_{5}]+\frac{1}{2}c^{22}m_{22}[s_{5}]
			\vspace{1mm} \\
			m_{12}[s_{5}+1]=m_{1}[s_{5}]
			\vspace{1mm} \\
			m_{22}[s_{5}+1]=m_{2}[s_{5}]
			\vspace{1mm} \\
			M_{1}[s_{4},s_{1}]=m_{1}[s_{5}+2]/f[s_{4}]
			\vspace{1mm} \\
			M_{2}[s_{4},s_{p}]=m_{2}[s_{5}+2]/f[s_{4}]
		\end{array}
	}
\end{align}
with, $2s_{1} \leq s_{5} \leq N+2s_{1}$ and $s_{p}=q-s_{1}$. The $f$, $b$ vectors are represent the factorials of the Eq. (\ref{eq:25_G0lE}), binomial coefficients of Eq. (\ref{eq:16_GQE}),
\begin{align}\label{eq:a7_bV}
	\boxed{
		b[s+1]=\left(\frac{q-s}{s+1}\right)b[s]
	}
\end{align}
respectively. Finally, in this section, what is left are just multiplying the defined vector forms of the functions,
\begin{align}\label{eq:a8_P1M1P2M2V}
	\boxed{
		\begin{array}{c}
			\left(PM\right)_{1}[s_{4},s_{1}]=b[s_{1}]P_{1}[s_{4},s_{1}]M_{1}[s_{4},s_{1}]
			\vspace{1mm} \\
			\left(PM\right)_{2}[s_{4},s_{1}]=b[s_{1}]P_{2}[s_{4},s_{1}]M_{2}[s_{4},s_{1}]
		\end{array}
	}
\end{align}
taking into account Eqs. (\ref{eq:23_G0AS}, \ref{eq:25_G0lE}), inserting the sum appropriately in the Eq. (\ref{eq:16_GQE}).
\section{\label{appc:Vec_form_mlc2} Vectorized form for the functions in Section \ref{sec:mlc2}}
\begin{align}\label{eq:b1_JBV}
	\boxed{
		B[s_{2},1]
		=br[s_{2}]
		B[s_{2}-1,1]
	}
\end{align}
\begin{align}\label{eq:b2_JBV2}
	\boxed{
		B[s_{2},s_{6}]
		=bc[s_{qe}]
		B[s_{2},s_{6}-1]
	}
\end{align}
\begin{align}\label{eq:b3_Jps1234}
	\boxed{
		\begin{array}{c}
			ps_{1}[s_{4},1]=psr_{1}[s_{4},1]ps_{1}[s_{4}-1,1]
			\vspace{1mm} \\
			ps_{1}[s_{4},2]=psr_{1}[s_{4},2]ps_{1}[s_{4}-1,2]
			\vspace{1mm} \\
			ps_{2}[s_{4},1]=psr_{2}[s_{4},1]ps_{2}[s_{4}-1,1]
			\vspace{1mm} \\
			ps_{2}[s_{4},2]=psr_{2}[s_{4},2]ps_{2}[s_{4}-1,2]
			\vspace{1mm} \\
			ps_{1}[s_{4},s_{6}]=psc_{1}[s_{4},s_{6}]ps_{1}[s_{4},s_{6}-1]
			\vspace{1mm} \\
			ps_{2}[s_{4},s_{6}]=psc_{2}[s,s_{6}]ps_{2}[s_{4},s_{6}-1]
		\end{array}
	}
\end{align}
Finally in vector forms for $m^{1}$ and $m^{2}$ functions we have,
\begin{align}\label{eq:b4_51Jm1m2V}
	\begin{array}{c}
		m_{1}[s_{4},1]=4r_{11}[s_{4},1]m_{1}[s_{4}-2,1]+r_{12}[s_{4},1]m_{1}[s_{4}-1,1]
		\vspace{1mm} \\
		m_{1}[s_{4},2]=4r_{11}[s_{4},2]m_{1}[s_{4}-2,2]+r_{12}[s_{4},2]m_{1}[s_{4}-1,2]
		\vspace{1mm} \\
		m_{2}[s_{4},1]=4r_{21}[s_{4},1]m_{2}[s_{4}-2,1]+r_{22}[s_{4},1]m_{2}[s_{4}-1,1]
		\vspace{1mm} \\
		m_{2}[s_{4},2]=4r_{21}[s_{4},2]m_{2}[s_{4}-2,2]+r_{22}[s_{4},2]m_{2}[s_{4}-1,2]
	\end{array}
\end{align}
\begin{align}\label{eq:b5_Jm1V}
	\boxed{
		\begin{array}{rl}
			m_{1}[s_{4},s_{6}]=\frac{1}{4}c_{11}[s_{4},s_{6}]m_{1}[s_{4},s_{6}-2] \\
			+c_{12}[s_{4},s_{6}]m_{1}[s_{4},s_{6}-1]
		\end{array}
	}
\end{align}
\begin{align}\label{eq:b6_Jm2V}
	\boxed{
		\begin{array}{rl}
			m_{2}[s_{4},s_{6}]=\frac{1}{4}c_{21}[s_{4},s_{6}]m_{2}[s_{4},s_{6}-2] \\
			+c_{22}[s_{4},s_{6}]m_{2}[s_{4},s_{6}-1]
		\end{array}
	}
\end{align}

\end{document}